# An efficient plate heater with uniform surface temperature engineered with effective thermal materials

Yichao Liu[1], Wei Jiang[1], Sailing He[1,2], and Yungui Ma[1,3,*]

[1]*State Key Lab of Modern Optical Instrumentation, Centre for Optical and Electromagnetic Research, Zhejiang University, Hangzhou 310058, China*
[2]*Department of Electromagnetic Engineering, School of Electrical Engineering, Royal Institute of Technology, S-100 44 Stockholm, Sweden*
[3]*Cyber Innovation Joint Research Center, Zhejiang University, Hangzhou 310058, China*

Corresponding e-mail: yungui@zju.edu.cn

**Extended from its electromagnetic counterpart, transformation thermodynamics applied to thermal conduction equations can map a virtual geometry into a physical thermal medium, realizing the manipulation of heat flux with almost arbitrarily desired diffusion paths, which provides unprecedented opportunities to create thermal devices unconceivable or deemed impossible before. In this work we employ this technique to design an efficient plate heater that can transiently achieve a large surface of uniform temperature powered by a small thermal source. As opposed to the traditional approach of relying on the deployment of a resistor network, our approach fully takes advantage of an advanced functional material system to guide the heat flux to achieve the desired temperature heating profile. A different set of material parameters for the transformed device has been developed, offering the parametric freedom for practical applications. As a proof of concept, the proposed devices are implemented with engineered thermal materials and show desired heating behaviors consistent with numerical simulations. Unique applications for these devices can be envisioned where stringent temperature uniformity and a compact heat source are both demanded.**

## 1. Introduction

The combination of transformation optics and metamaterials has allowed enormous powers to manipulate electromagnetic (EM) waves and configure artificial devices with functionalities deemed impossible before, as typically represented by invisible cloaks and perfect lenses.[1,2] The invariance of Maxwell's equations under coordinate transformation is the physical origin establishing the transformation of a virtual geometry into a concrete physical device with desired functions. In addition to cloaking, other interesting EM phenomena or instrumental applications such as wormholes, omni-directional retroreflectors, wave rotators and beam-modulators have been proposed applying these techniques.[3-6] While the majority of research in this direction is still focusing on improving the practicality of EM cloaks,[7-10] the same technique associated with metamaterials has been broadly extended to other partial differential equations governing evolutions of different physical quantities, such as acoustic waves, flux, matter waves or even Schrödinger waves.[11-14] The precise and desirable manipulation of the propagation of these waves has led to many conceptual breakthroughs and technical advancements for these fields.

In recent years, transformation thermodynamics and manipulation of heat flux by pre-defined diffusion paths have attracted great attentions: steering heat flux to realize a thermal cloak, in particular.[15,16] Initially in the literature, these were discussed by applying a transformation of the static thermal conduction equation, $\nabla \kappa \nabla u = 0$, where $\kappa$ and $u$ represent thermal conductivity and the temperature field, respectively, thus giving rise to thermostatic artificial devices such as thermal cloaks, concentrators or rotators reproducing their EM counterparts.[15a] In 2012[16a], Guenneau et al. extended this operation on the transient thermal conduction equation, $\rho c \cdot \partial u/\partial t = \nabla \cdot (\kappa \nabla u)$, where $\rho$ and $c$ are the matter density and thermal capacity, respectively. They proposed similar functional devices but with transient responses that better appeal to many practical applications such as thermal shielding or harvesting.[17] The idea of transient heat cloaking was strictly implemented via the precise manufacture of artificial thermal materials, independently by the current authors and Wegener's group.[16b, 16c] Quite recently, two Singaporean groups reported a simplified method to fabricate similar cloaking devices utilizing a bilayer-material approach originally developed for a static magnetic field cloak.[15c, 15d, 18]

In fact we need to point out that a thermal cloak works on the heat flux by steering it around an obstacle, but the temperature field in the hidden region actually rises up when the environment changes, obeying the second thermodynamic theorem. Elegant periodical nanostructures rendering phononic bandgaps for acoustic waves carrying heat may help to finally prohibit atomic vibration and produce a perfect thermal cloak.[19] Nonetheless, the unique capability to design a desirable flux path provides numerous opportunities to tailor the heat to improve existing thermal-related technologies and conceive new ones, which is further supported by the fact that the transformed thermal devices can be more easily fabricated compared with the transformed EM devices.[16b]

In this work we propose another instrumental application of transformation thermodynamics to design an efficient plate heater that can transiently provide a large and homogeneous-temperature surface powered by a very small thermal source. In contrast to a convectional heater baked by a large source, we employ a specifically engineered conduction plate consisting of complex thermal materials to guide and spread the heat flux. To our best knowledge, this will be the first attempt to configure the heating temperature field through material deployment. The proof-of-concept devices, designed with a feeding source covering only one percent of the output temperature surface, are implemented with precisely engineered thermal materials exhibiting prominent heating functionalities consistent with numerical predictions. Potential applications are envisioned, e.g., of an on-chip thermal modulation, where temperature uniformity and compact source are both demanded.[20]

## 2. Results and Discussions

### 2.1 Design Algorithm

A plate heater is a very common device that provides a specific temperature environment for a physical or chemical reaction. As schematically shown in **Figure 1**a, a conventional heater consists of two general parts: a conduction plate and a thermal source. To obtain a large and transiently homogeneous usable temperature surface, the practical way is, on the heating source, to deploy the resistor wires to form a complex network covering an area as large as possible. The proper layout can meet the requirement of most free-space heating applications. However it may not be suitable for some specific *in-situ* occasions such as on-chip lasing or involving weak signal measurements where less external electric or current involvement with a compact source is highly desired in addition to a uniform heating surface. [20] Our contribution here is to provide an alternative approach to meet these stringent requirements using transformation thermodynamics. A complex functional conduction plate is configured under the powering of a small source transiently providing a homogeneous-temperature output surface.

Our design algorithm is schematically shown in **Figure 1**b by the two-dimensional carton pictures mapping a virtual small cylindrical object of radius $b$ and constant height $h$ onto a flat physical plate to form a circular truncated cone, by linearly enlarging the horizontal ($x$-$y$ plane) radius from bottom $b$ to surface $a$ ($\gg b$). In the cylindrical coordinate this mapping from the virtual ($r$, $\phi$, $z$) to physical space ($r'$, $\phi'$, $z'$) can be written by

$$r' = \gamma \cdot r, \quad \phi' = \phi, \quad z' = z \qquad (1)$$

where $\gamma = (a-b)z/(bh)+1$ and the bottom face is at the $z = 0$ plane. In the virtual space, the small heater adapted from a traditional design will provide a uniform temperature surface with a working surface area nearly equal to the heating source, i.e., $\pi b^2$. In an ideal case the flux trajectories in the transformed plate will be spread and guided to produce a larger output temperature surface ($\pi a^2$) powered by the same-sized source. Mathematically similar to an optical transformation building EM beam transformers, [21] this spatial operation modifies the flux beam size by creating a thermal expander or concentrator dependent on the boundary temperature difference between the ends. The variation time of the flux waist is equal to the ratio $a/b$, which also determines the anisotropy of the device and the practical feasibility. In this work we only consider it as a transient flux expander in order to obtain a heat conduction plate.

To acquire the dynamic response, we start with the transformation from the transient thermal conduction equation and also take an approximation to adopt the rescaled form as previously needed in our transient cloaking experiment, [16b]

$$\frac{\partial u}{\partial t} = \nabla' \cdot (\kappa'/\rho' c' \nabla' u), \qquad (2)$$

where $\rho'c' = \rho c/\det(\boldsymbol{J})$, $\boldsymbol{\kappa}' = \boldsymbol{J}\boldsymbol{\kappa}\boldsymbol{J}^{\mathrm{T}}/\det(\boldsymbol{J})$ and $\boldsymbol{J}$ is the Jacobean matrix defining the coordinate mapping. **Equation 2** is valid only at the condition $\nabla'(1/\rho'c') \approx 0$. In the previous cloaking experiment, this condition was satisfied with properly selected combinations of multiple ingredient materials. [16b] In the present case it is practically satisfied by assuming a variable background $\rho c$ cancelling the change of $\det(\boldsymbol{J})$ across the sub-layers. Compared with a homogeneous background, this assumption will not influence the heat flux trajectories as an ideal transformer but does change the final temperature values on the output surface. With this approximation, the implementation process is greatly simplified and only needs to take into account the anisotropic thermal diffusivity $\boldsymbol{\alpha}' (= \boldsymbol{\kappa}'/\rho'c')$. To do this we assume a virtual plate consisting of ten sub-layers and transform each sub-layer separately by the mapping $r'_n = \gamma_n \cdot r_n$ where $\gamma_n = na/(10b)$ and $n$ is the number of layers. After a simple algebraic process the anisotropic diffusivity for each sub-layer reads the form

$$\boldsymbol{\alpha}'_n = \mathrm{diagonal}(\gamma_n, \gamma_n, \gamma_n^{-1})\alpha_{n0}, \qquad (3)$$

where $\alpha_{n0}$ is the background diffusivity. In the above transformation, ultrathin transition layers with off-diagonal parameters between two sub-layers are practically neglected in our modeling and implementation. This approximation will cause a slight influence on the flux characteristics of the device under the adiabatic change of the sub-layer radius as satisfied in our design and experiment.

## 2.2 Simulations and Implementation

Here in the proof-of-concept experiment, we use structural parameters $a = 10b = 50$ mm and $h = 5$ mm to engineer a thermal plate with a feeding source surface 100 times smaller than the output temperature plane. Two experiments are conducted following different transformation schemes. The first one strictly satisfies the transformed parametric profile defined by **Equation 3** and the second one uses a homogeneous material profile but with a large anisotropy following another transformation mapped from a truncated cone. Firstly we perform the numerical examination on the device defined by **Equation 3** by COMSOL and also a control made of pure aluminum, which has the same flat truncated cone topology as the sample and also the background diffusivity (90.6 mm$^2$/s). In simulation we use a 5W-heater source and a convection boundary condition with the surrounding air (of coefficient 15 W·m$^{-2}$K$^{-1}$). The loss due to heat radiation is neglected as we work at moderate temperatures less than 100 °C.

**Figures 2**a-**2**f show the snapshots of the output temperature profile at various heating times for the sample and the control, respectively. The inhomogeneous temperature rising across the surface is clearly observed for the control while the transformed sample shows a highly uniform temperature field at different times. **Figure 2**g plots the evolution of the temperature field taken at the center and one edge point of the circular heating surface for both cases. In contrast with the control, the two temperature rising curves of the sample superimpose over each other over the observing time window. Their difference ratio, defined by $(T_c-T_e)/T_c$ for the control, remains at about 3% and almost zero for the transformed sample, confirming the validity of the designation algorithm. Note that the existence of boundary convection with the surrounding air leads to the nonlinear increment of temperature with respect to the heating time but does not influence the temperature distribution feature on the output surface, which guarantees practical application.

The 10-layered inhomogeneous conduction plate is implemented using the proper combinations of 0.1 mm thick aluminum and artificially synthesized graphite sheets. The material parameters including those used in the second experiment can be found in our previous paper on cloaking, wherein **Equation 5** describes the averaging formula for the effective thermal parameters.[16b] **Figure 3**a plots the in-plane (//) and out-of-plane ($\perp$) diffusivities of different sub-layers. The calculated values denoted by symbols are obtained according to the real structure of the sample, and the theoretical ones denoted by solid lines are derived from **Equation 3**. They agree well except for some points on the in-plane diffusivity due to the limit of finite ingredient thickness. The tightly stacked sample is finally placed in a shallow Teflon container, which has a topology complementary to the sample, with the top surface covered by a 50 μm kapton. The small conductivity (0.25 W·m$^{-1}$·K$^{-1}$) of Teflon is expected to induce little effect on the thermal behavior of the device. **Figure 3**b schematically shows the sample device and the measurement setup. The carton picture of the sample is intentionally enlarged along the $z$-axis to allow clear viewing. In the experiment, a 1.5 W-heater is glued on the bottom of the device as the source and the evolution of the temperature field on the top output surface is captured through a Mikron 7500l infrared camera through thermal emission.

**Figures 4**a-**4**f give the measured snapshots of the temperature field for the aluminum control and the sample device at three different heating times. The transient evolution of the temperature for the sample is animated in the supporting **Figure S1**. Compared with the control, the sample device elaborates a transient and ultra-homogeneous temperature field across the entire output surface. In **Figure 4**g this characteristic is clearly manifested from the evolution curves of the temperature field respectively taken at the center and one edge point for both cases. Their difference ratio of $(T_c-T_e)/T_c$ as plotted in the inset of **Figure 4**g remains at about 2.75% for the control and almost zero for the sample. These trends agree with the simulation results very well. In addition, the imperfection of the sample fabrication, in particular the existence of some unwanted inter-sheet gaps, could induce additional interfacial thermal resistance and lead to the reduction of the rising temperature slope compared with the simulation. Apparently this imperfection has a slight influence on the desired uniform heating-up characteristic. Practically this defect can be minimized with tight glue bonding or by applying high-pressure force in measurements.

The first samples are transformed from a small cylindrical virtual plate, as shown in **Figure 1**b, by gradually stretching the radius from the bottom at a constant height. In practice the overall height of the plate can be rather small to produce a thin conduction plate. On the other hand, a device of similar function can be conceived by compressing a very tall virtual truncated cone (height $h$) into a low one (height $h' << h$) with the transverse topology invariant. Such a tall virtual plate with a height much larger than the average radius will also give a homogeneous-output-temperature surface (of area $\pi a^2$). This linear compressive operation will transform the original isotropic tall plate into a shallow one composed of a large uniaxial anisotropy: $\alpha_{//}/\alpha_{\perp} = (h/h')^2 \gg 1$. Here we implement it with the same structural parameters as the first inhomogeneous one: $a = 10b = 10h' = 50$ mm, and at $h = 14h'$. Three ingredient materials of copper, iron and kapton with a respective thickness ratio of 0.5:0.4:0.1 are combined together to satisfy the anisotropic diffusivities $\alpha_{//} = 67.81$ mm$^2$/s and $\alpha_{\perp} = 0.36$ mm$^2$/s.

**Figures 5**a-**5**f show the simulated and measured temperature snapshots at three different heating times for the anisotropic plate heater. An animation in the supporting **Figure S2** is provided to show the transient temperature variation for the measured results. From these pictures the homogeneous heating-up characteristic for the device is clearly evidenced both numerically and experimentally. **Figure 5**g plots the rising temperature curves corresponding to the center and one edge point of the heating plate. The temperature difference between these two points is negligibly small. The rising-slope difference between the simulation and experiment arises from the fabrication defect and also possibly the environmental difference between the simulation and experiment. It is noted that the second device heats up the temperature much more slowly than the first inhomogeneous one as shown in **Figure 4**. This is mainly due to the practical limitation that large anisotropy for the second device is only fulfilled at the cost of using a very small background thermal diffusivity (5 mm$^2$/s) for the second sample.

## 3. Conclusions

In this work we have utilized the transformation technique to design an efficient heating device able to transiently control heat flux and provide a large and uniform-temperature surface powered by a small source. The virtual object associated with a specific transformation is properly combined to offer different material designs of the same thermal functions. Between the two categories of composition proposed here, the inhomogeneous conduction plate produces a gradual change in their constitutive parameters and supports a fast rise in temperature, possessing more practical potentials, while the homogeneous one is relatively easier for fabrication. Unique applications for the proposed devices may be found in some special heating occasions where temperature uniformity and a small power source (or little electric involvement) are both demanded such as for a temperature-mediated on-chip laser or an *in-situ* heating transmission electron microscopy.[20,22] The stacking approach developed to engineer the thermal devices has very high practical feasibility and can be accurately implemented by a standard technique such as sputtering deposition for these specific applications. The current research is another good example of the powerful application of coordinate transformation on wave/flux manipulation and may help to broaden research in acquiring extraordinary ways to control and utilize heat energy or configure novel heat devices.

**Acknowledgements**

The authors are grateful to the partial supports from NSFCs 61271085, 60990322 and 91130004, the National High Technology Research and Development Program (863 Program) of China (No. 2012AA030402), the Program of Zhejiang Leading Team of Science and Technology Innovation, NCET, MOE SRFDP of China, and Swedish VR grant (# 621-2011-4620) and SOARD and the support by the Fundamental Research Funds for the Central Universities.

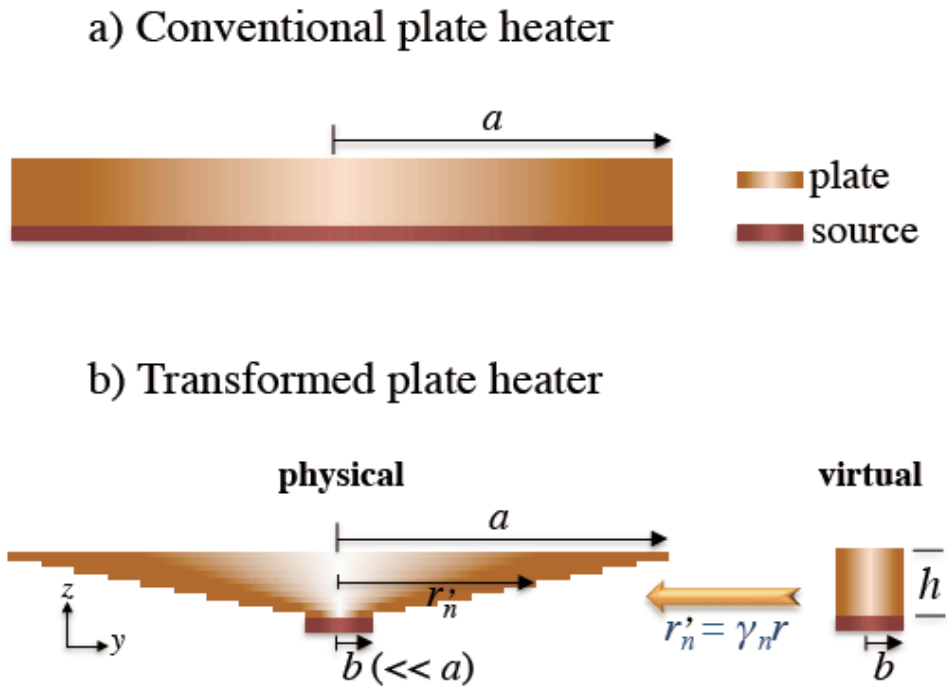

**Figure 1** Schematic of the transformation process. (a) A conventional plate heater and (b) a transformed plate heater. In (b), a virtual cylindrical object of radius $b$ and height $h$ is stretched into a truncated cone (with bottom radius $b$ and top radius $a$) under a constant height by following the formula $r'_n = \gamma_n \cdot r_n$. Compared with the implemented parameters the sample height is intentionally enlarged to provide clear viewing.

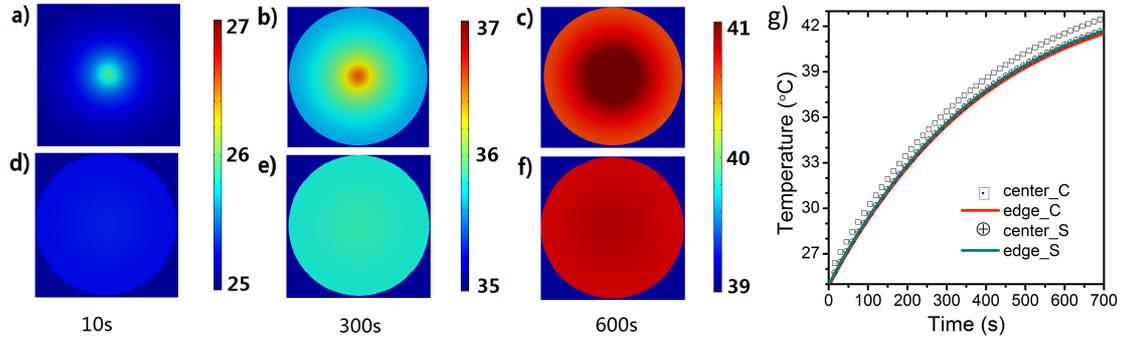

**Figure 2** Simulated heating characteristic of the transformed inhomogeneous plate heater. (a)-(c) temperature snapshots of the aluminum control, (d)-(f) temperature snapshots of the sample device, and (g) the rising temperature curves taken at the center and one edge point of the circular output surface for the control (R) and the sample (S).

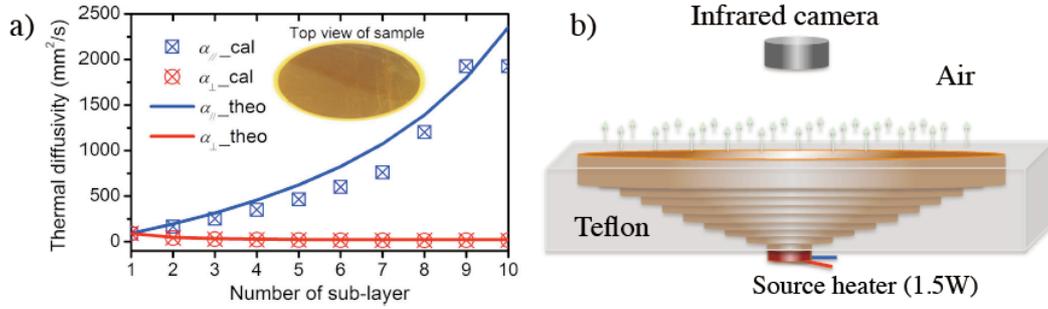

**Figure 3** Device parameters and measurement setup. (a) In-plane (//) and out-of-plane (⊥) thermal diffusivity values calculated according to the real sample structure (symbols) and theoretical ones (solid lines) defined by **Equation 3** for different sub-layers. (b) Measurement setup. The inset in (a) gives a top view of the implemented device. In (b), the sample is placed inside a Teflon container and powered by a 1.5 W source underneath. An IR camera is used to capture the temperature field through thermal emission.

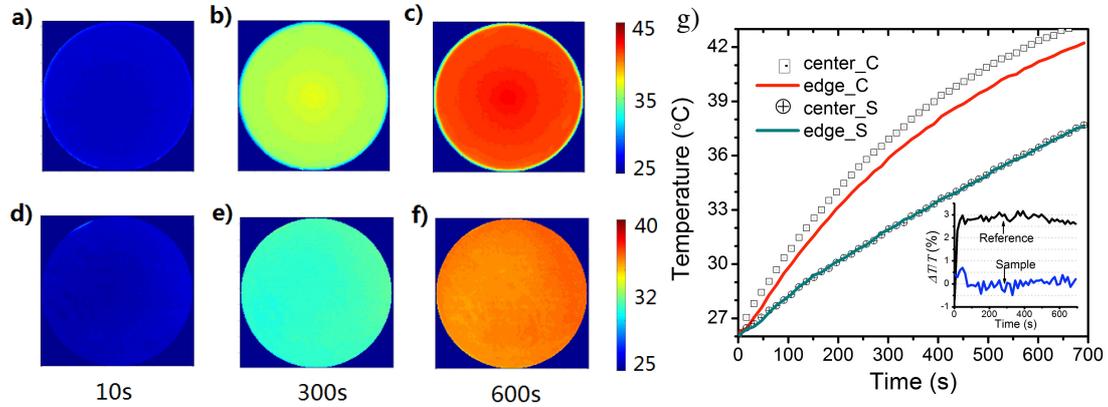

**Figure 4** Measured heating characteristics of the transformed inhomogeneous plate heater. (a)-(c) Temperature snapshots of the aluminum control, (d)-(f) temperature snapshots of the sample device and (g) the rising temperature curves taken at the center and one edge point of the circular output surface for the control (R) and the sample (S). The inset in (g) plots the temperature difference ratio, $(T_c-T_e)/T_c$, between these two points.
13

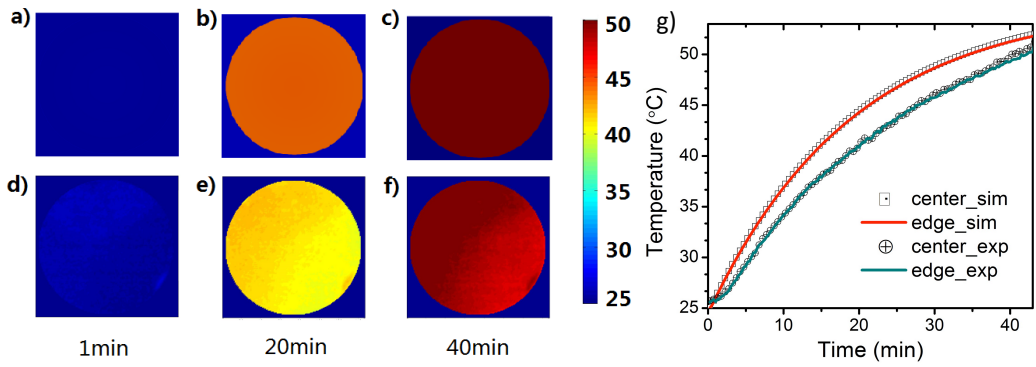

**Figure 5** Heating characteristic of the transformed homogeneous plate heater. (a)-(c) Simulated temperature snapshots, (d)-(f) measured temperature snapshots and (g) the simulated and measured rising temperature curves taken at the center and one edge point of the circular output surface.